\documentclass[10pt]{article}
\usepackage{graphicx}
\usepackage{lineno}

\def\Title#1{\begin{center} {\Large #1 } \end{center}}
\def\Author#1{\begin{center}{ \sc #1} \end{center}}
\def\Address#1{\begin{center}{ \it #1} \end{center}}

\newcommand\pubblock{\rightline{\begin{tabular}{l} Proceedings of the Fifth Annual LHCP\\ \pubnumber\\
         \pubdate  \end{tabular}}}

\newenvironment{Abstract}{\begin{quotation} \begin{center} 
             \large ABSTRACT \end{center}\bigskip 
      \begin{center}\begin{large}}{\end{large}\end{center} \end{quotation}}

\newenvironment{Presented}{\begin{quotation} \begin{center} 
             PRESENTED AT\end{center}\bigskip 
      \begin{center}\begin{large}}{\end{large}\end{center} \end{quotation}}

\def\beq{\begin{equation}}
\def\eeq#1{\label{#1}\end{equation}}
\def\eeqn{\end{equation}}

\def\beqa{\begin{eqnarray}}
\def\eeqa#1{\label{#1}\end{eqnarray}}
\def\eeqan{\end{eqnarray}}

\let\bar=\overbar

\def\Dslash{\not{\hbox{\kern-4pt $D$}}}
\def\dslash{\not{\hbox{\kern-2pt $\del$}}}

\def\msb{{\bar{\ssstyle M \kern -1pt S}}}

\usepackage{color}

\newcommand\pubnumber{ ATL-PHYS-PROC-2017-209 }

\newcommand\pubdate{\today}

\def\affiliation{
On behalf of the ATLAS Collaboration, \\
Department of Physics \\
University of Oxford,  Parks Road, OX1 3PU Oxford, United Kingdom}

\begin{document}
\large
\begin{titlepage}
\pubblock

\vfill
\Title{ Higgs boson measurements in the $WW^{\ast}$, $\tau\tau$, and $\mu\mu$ channels with the ATLAS Experiment  }
\vfill

\Author{ KATHRIN BECKER  }
\Address{\affiliation}
\vfill
\begin{Abstract}

The status of Higgs boson measurements in the $WW^{\ast}$, $\tau\tau$, and $\mu\mu$ decay modes with the ATLAS experiment is presented. These measurements are done using $pp$ collision data from the Large Hadron Collider with either the full Run 1 dataset at $\sqrt{s}= 7\,\mathrm{TeV}$ and $\sqrt{s}= 8\,\mathrm{TeV}$ or with a partial Run 2 dataset at $\sqrt{s}= 13\,\mathrm{TeV}$ of either $5.81\,\mathrm{fb}^{-1}$ or $36.1\,\mathrm{fb}^{-1}$. 
The different couplings to the Higgs boson are probed by measuring the production cross sections and extracting the signal strengths. Depending on the analysis, the measurements are sensitive to the gluon-fusion or vector-boson fusion Higgs boson production, or associated production of a vector boson and a Higgs boson. 
\end{Abstract}
\vfill

\begin{Presented}
The Fifth Annual Conference\\
 on Large Hadron Collider Physics \\
Shanghai Jiao Tong University, Shanghai, China\\ 
May 15-20, 2017
\end{Presented}
\vfill
\end{titlepage}
\def\thefootnote{\fnsymbol{footnote}}
\setcounter{footnote}{0}
%

\normalsize 


\section{Introduction}

In this document the status of measurements of the Standard Model~(SM) Higgs boson in the $WW^{\ast}$, $\tau\tau$, and $\mu\mu$ decay modes are discussed which are done with the ATLAS experiment~\cite{Aad:2008zzm} at the Large Hadron Collider~(LHC). 
These measurements are performed using $pp$ collision data from the Large Hadron Collider with either the full Run 1 dataset at $\sqrt{s}= 7\,\mathrm{TeV}$ and $\sqrt{s}= 8\,\mathrm{TeV}$ or a partial Run 2 dataset at $\sqrt{s}= 13\,\mathrm{TeV}$. 

All measurements presented aim to probe the couplings of the SM Higgs boson to either bosons ($W$) or fermions ($\mu$ and $\tau$). This is done by measuring the Higgs production cross section times branching fraction, 
$\sigma_{i\rightarrow f} = \sigma_{i} \cdot \mathrm{B}_{f}$ for a specific production process $i$ and a decay mode $f$. The signal strength, $\mu$, is defined as the ratio of the measured Higgs boson rate to its SM prediction:
\begin{equation}
\mu_{i\rightarrow f} = \frac{\sigma_{i} \cdot \mathrm{B}_{f}}{(\sigma_{i})_{\mathrm{SM}} \cdot (\mathrm{B}_{f})_{\mathrm{SM}}}.
\end{equation}
The subscript ``SM'' refers to the respective SM prediction.

Each analysis (channel) is sensitive to different Higgs boson production processes. In all presented decay modes, Higgs boson production via gluon-gluon fusion~(ggF) and via vector-boson fusion~(VBF) are probed. The associated $VH$ production is only discussed for $H \rightarrow WW^{\ast}$ where a Run~2 analysis is available. 

\section{Measurements in $H \rightarrow WW^{\ast}$}

In Run~1 a complete set of analyses has been performed in the $H \rightarrow WW^{\ast}$ decay mode, investigating the ggF, VBF, and $VH$ production processes. The combined ggF and VBF analysis has led to the observation of the $H \rightarrow WW^{\ast}$ decay mode with a significance of $6.1\sigma$, measuring a signal strength of $\mu_{\mathrm{ggF+VBF}} = 1.09\,^{+0.16}_{-0.15}\,\mathrm{(stat.)}\;^{+0.17}_{-0.14}\,\mathrm{(syst.)}$~\cite{ATLAS:2014aga}. Also evidence was established for the VBF production process. It can be seen, that already in Run~1 the combined signal strength measurement has statistical and systematic uncertainties of a similar size. 
In addition to the signal strength and inclusive cross section measurements, first measurements of fiducial and differential cross sections have been performed for the ggF production process~\cite{Aad:2016lvc}. A search for $VH$ production~\cite{Aad:2015ona} has yielded a signal strength of $\mu_{VH} = 3.0\,^{+1.3}_{-1.1}\,\mathrm{(stat.)}\;^{+1.0}_{-0.7}\,\mathrm{(syst.)}$ corresponding to an observed significance of $2.5\sigma$.

At $13\,\mathrm{TeV}$ a first analysis investigating the VBF and $WH$ production mode has been performed with a dataset of $5.8\,\mathrm{fb}^{-1}$~\cite{HWWConf}. To measure the VBF production cross section, two opposite-flavour ($e\mu$), opposite-sign leptons with $p_{\mathrm{T}}> 15\,\mathrm{GeV}$ are selected where the leading electron (muon) is required to have $p_{\mathrm{T}}> 25\,\mathrm{GeV}$ ($p_{\mathrm{T}}> 22\,\mathrm{GeV}$) and the mass of the dilepton system is required to be larger than $10\,\mathrm{GeV}$.
A requirement of $N_{\mathrm{jet}}\,{\ge}\,2$ is applied to enhance the contribution from VBF signal events.
Here, the jets are required to have $p_{\mathrm{T}}\,{>}\,25\,\mathrm{GeV}$ if $|\,\eta\,|\,{<}\,2.4$ and $p_{\mathrm{T}}\,{>}\,30\,\mathrm{GeV}$ for $2.4\,{<}\,|\,\eta\,|\,{<}\,4.5$.
A $b$-jet veto is applied to reject top-quark background and a requirement on the invariant mass of the $\tau\tau$ system ($m_{\tau\tau}$), $m_{\tau\tau} < 66.2\,\mathrm{GeV}$, is applied to reject background from $Z\rightarrow\tau\tau$. An ``outside-lepton-veto'' (OLV) is applied which requires the leptons to reside within the rapidity gap spanned by the two leading jets, as preferred by VBF event topology.
The ``central-jet-veto'' (CJV) rejects events with additional jets with $p_{\mathrm{T}} > 20\,\mathrm{GeV}$ in the rapidity gap of the two leading jets, suppressing backgrounds with more jet activity, such as $t\bar{t}$ production. To further improve the signal-to-background ratio a boosted decision tree~(BDT)~\cite{cart84,FREUND1997119,Friedman2002367} is trained using eight discriminating variables.
These variables exploit either properties of the $H \rightarrow WW^{\ast}$ decay, which is the case for $\Delta\phi_{\ell\ell}$, $m_{\ell\ell}$, and $m_{\mathrm{T}}$,
or properties of the VBF signature, which is the case for $\Delta y_{jj}$, $m_{jj}$, $p_{\mathrm{T}}^{\mathrm{tot}}$, $\sum_{\ell,j} m_{\ell j}$, and $\eta_\ell^{\mathrm{centrality}}$. The BDT score ranging from $-$1 to 1 is used to define the signal region by requiring the score to be larger than $-$0.8. Two bins of the remaining BDT score are defined with bin $[-0.8, 0.7]$ denoted SR1 and bin $[0.7, 1]$ denoted SR2.

\begin{figure}[tb]
\centering
\includegraphics[height=2.3in]{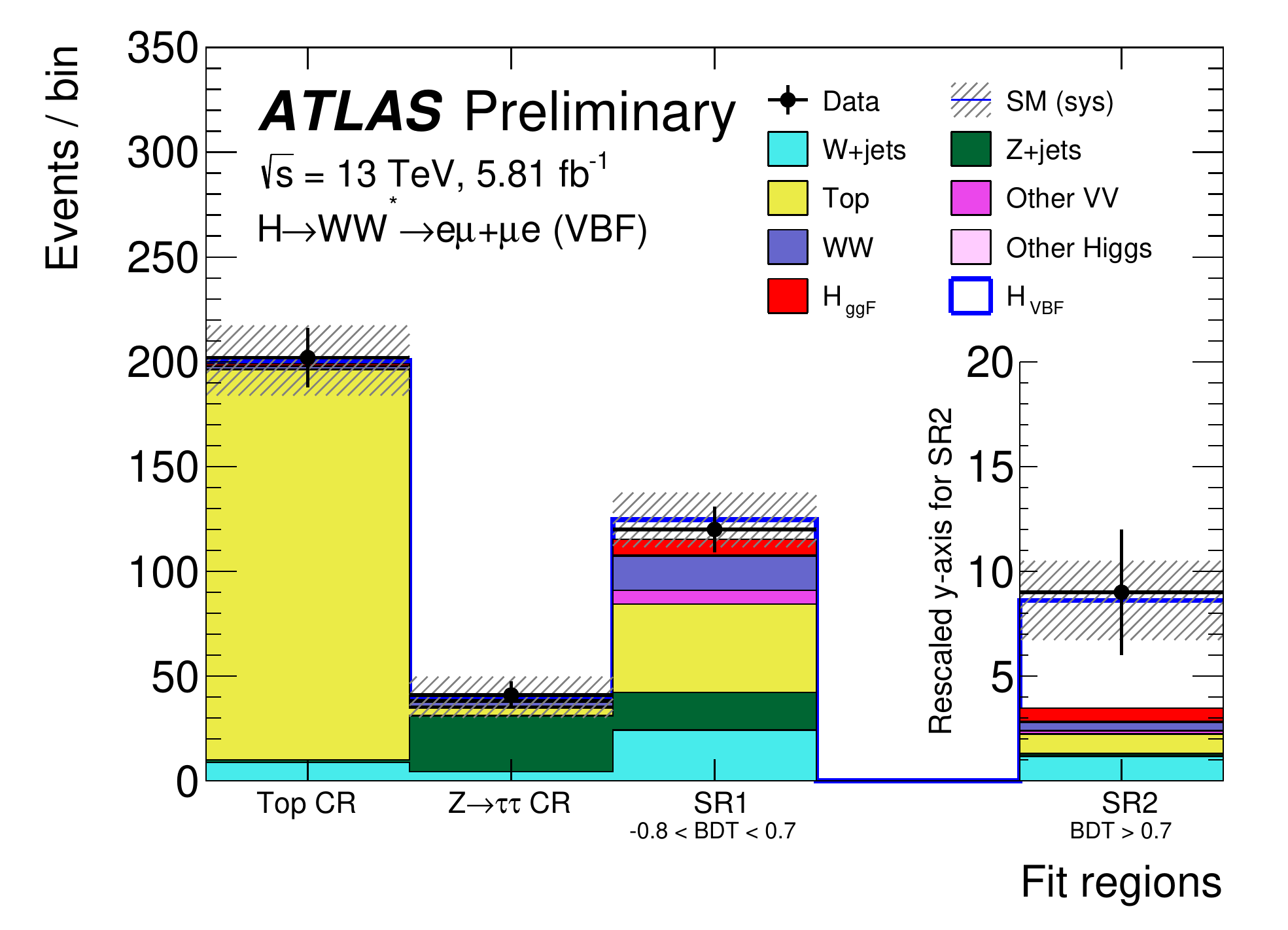}
\includegraphics[height=2.3in]{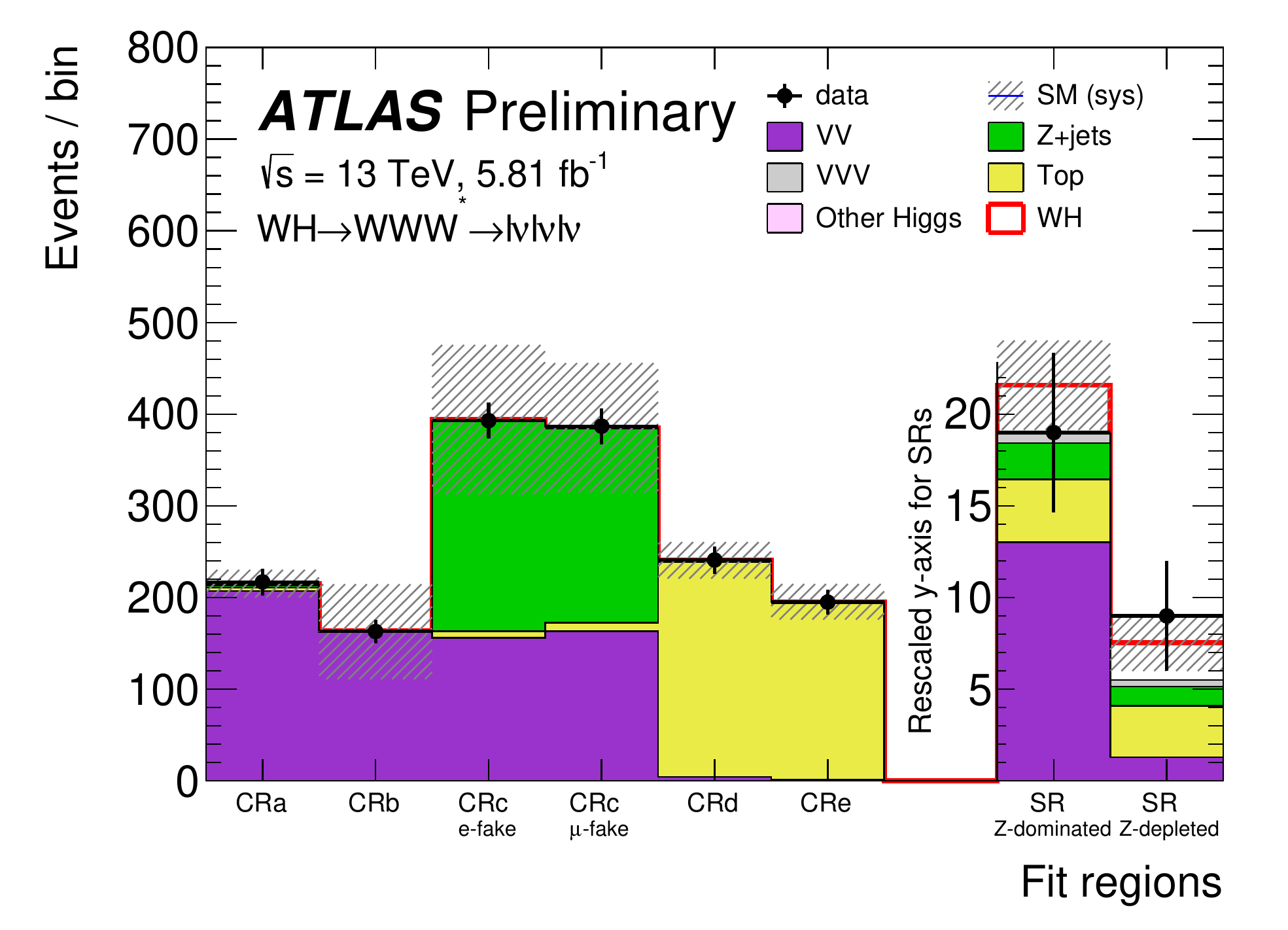}
\caption{ Fit regions of (a) the VBF analysis and (b) the $WH$ analysis~\cite{HWWConf}. The signal and background predictions are normalised to the results of the corresponding likelihood fit. The hatched band (denoted as ÒSM (sys)Ó) includes MC statistical, experimental, and theory systematic uncertainties associated with the prediction of the signal and background processes.}
\label{fig:ww}
\end{figure}

The largest background contributions in the SR originate from top-quark production, $WW$ production, and $Z\rightarrow\tau\tau$ production. $WW$ production is normalised to theory prediction, but the top-quark and $Z\rightarrow\tau\tau$ backgrounds are normalised to the data yield of a dedicated control region. The top-quark control region is defined by inverting the $b$-jet veto to a $b$-jet tag, while the $Z\rightarrow\tau\tau$ control region is defined by changing the $m_{\tau\tau}$ requirement to be within $25\,\mathrm{GeV}$ of $m_{Z}$. The VBF cross section and signal strength parameter are extracted by performing a simultaneous likelihood fit to SR1 and SR2 as well as the top-quark and $Z\rightarrow\tau\tau$ control regions. The fit regions are shown in Figure~\ref{fig:ww}(a) where signal and background predictions have been normalised to the results of the fit. The resulting VBF signal strength is 
$\mu_{\mathrm{VBF}} = 1.7^{+1.0}_{-0.8}\mathrm{(stat.)}^{+0.6}_{-0.4} \mathrm{(syst.)}$, which yields a cross section of 
$\sigma_{\mathrm{VBF}}\cdot \mathcal{B}_{H \rightarrow WW^{\ast}} = 1.4^{+0.8}_{-0.6}\mathrm{(stat.)}^{+0.5}_{-0.4} \mathrm{(syst.)}\; \mathrm{pb}$.
The observed (expected) significance is 
$1.9 \sigma$ ($1.2 \sigma$) assuming a Higgs boson mass of $125\,\mathrm{GeV}$. This first measurement of VBF production is still dominated by statistical uncertainties.

For the measurement of $WH$ production three isolated leptons with $p_{\mathrm{T}}> 15\,\mathrm{GeV}$ and a total electric charge of $\pm\,1$ are selected.  A requirement of $N_{\mathrm{jet}}\,{\le}\,1$ and a $b$-jet veto are applied to reject top-quark background. To reduce the $WZ/W\gamma^{\ast}$ background, the largest opposite-sign di-lepton invariant mass is required to be less than $200 \,\mathrm{GeV}$. Also a veto on $Z$ boson production is applied. To select Higgs boson event topology, the angular separation between lepton with unique charge and the closest lepton to it, $\Delta R_{\ell_0,\ell_1}$, is required to be smaller than 2 radians.
Then, two signal region categories are defined by the number of same-flavour opposite-sign (SFOS) lepton pairs. Events with at least one pair of SFOS leptons are classified in the ``$Z$-dominated'' category where background from $WZ/W\gamma^{\ast}$ and $Z$ boson production dominates,
while events without any SFOS pairs are classified in the ``$Z$-depleted'' category where background from top-quark production dominates.

Data control regions are defined for both signal regions to normalise top-quark, $WZ/W\gamma^{\ast}$, and $Z$ boson production to data yields. These six control regions and the two signal regions are fitted simultaneously to extract the $WH$ signal strength and production cross section. An overview on the fit regions is given in Figure~\ref{fig:ww}(b). The observed (expected) significance is 
$0.77 \sigma$ ($0.24 \sigma$) assuming a Higgs boson mass of $125\,\mathrm{GeV}$. The resulting signal strength is $\mu_{WH} = 3.2 ^{+3.7}_{-3.2}\mathrm{(stat.)} ^{+2.3}_{-2.7}\mathrm{(syst.)}$ yielding a cross sections of $\sigma_{WH}\cdot \mathcal{B}_{H \rightarrow WW^{\ast}} = 0.9 ^{+1.1}_{-0.9}\mathrm{(stat.)} ^{+0.7}_{-0.8}\mathrm{(syst.)}\; \mathrm{pb}$. The presented measurement is dominated by statistical uncertainties.

\section{Measurements in $H \rightarrow \tau\tau$}

\begin{figure}[tb]
\centering
\includegraphics[height=2.5in]{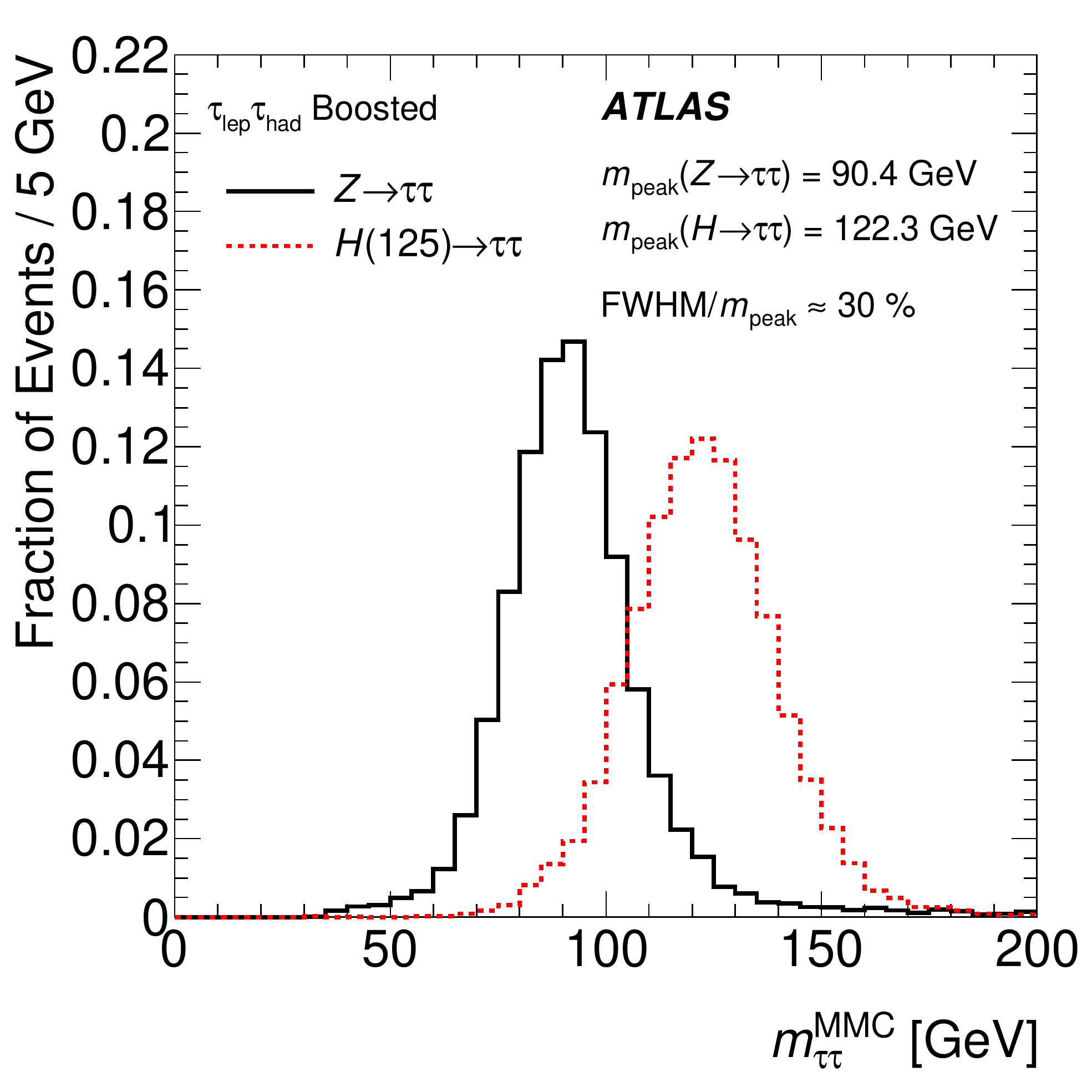}
\hspace{20mm}
\includegraphics[height=2.5in]{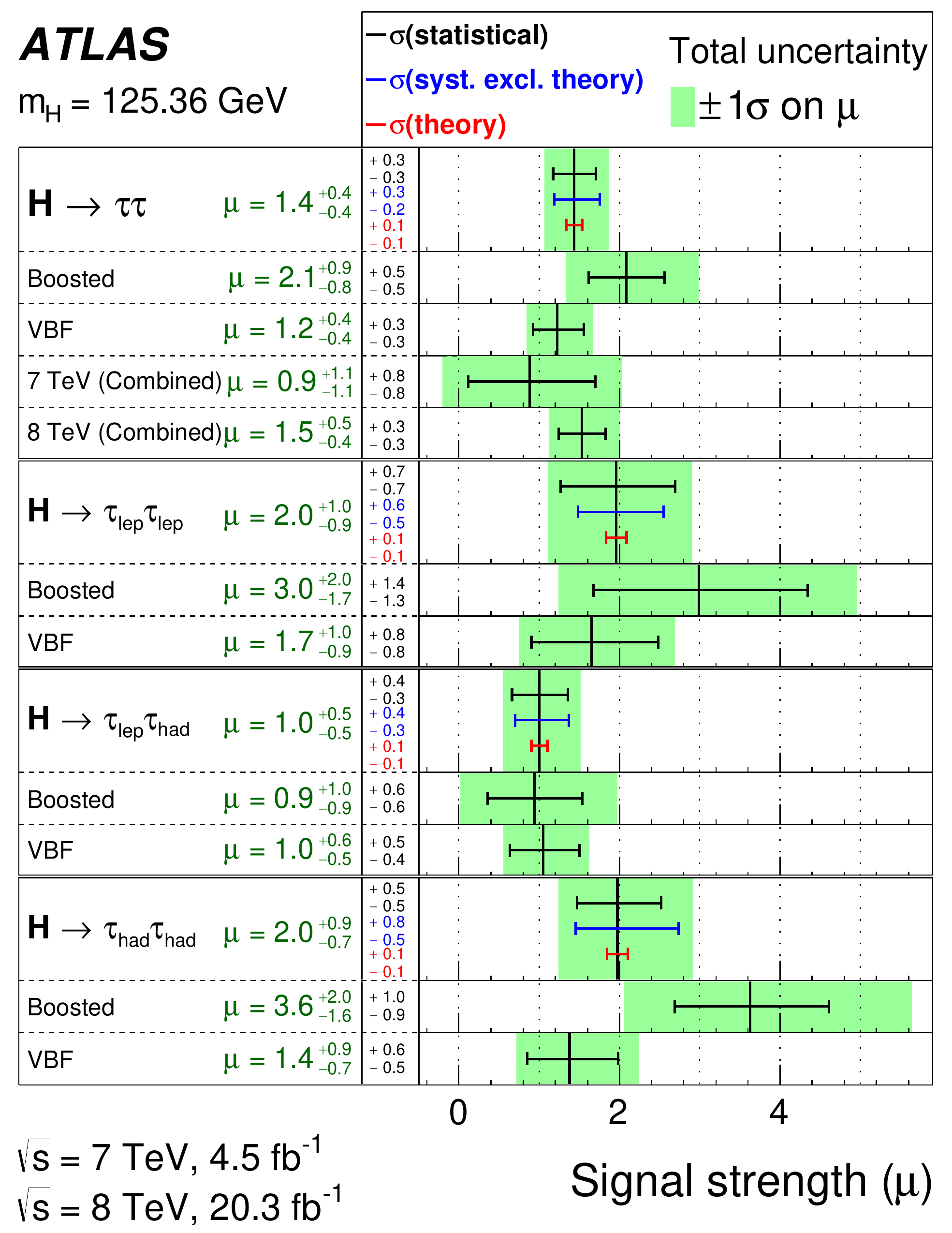}
\caption{ Fig.~(a) shows the reconstructed invariant $\tau\tau$ mass, $m_{\tau\tau}^{\mathrm{MMC}}$ for $H \rightarrow \tau\tau$ and $Z \rightarrow \tau\tau$ events in MC simulation and embedding respectively, for events passing the boosted category selection in the $\tau_{\mathrm{lep}}\tau_{\mathrm{had}}$ channel~\cite{Aad:2015vsa}. Fig.~(b) shows the best-fit value for the signal strength $\mu$ in the individual channels and their combination for the full ATLAS datasets at $\sqrt{s}= 7\,\mathrm{TeV}$ and $\sqrt{s}= 8\,\mathrm{TeV}$~\cite{Aad:2015vsa}. The total $\pm\,1\sigma$ uncertainty is indicated by the shaded green band, with the individual contributions from the statistical uncertainty (top, black), the experimental systematic uncertainty (middle, blue), and the theory uncertainty (bottom, red) on the signal cross section (from QCD scale, PDF, and branching ratios) shown by the error bars and printed in the central column.}
\label{fig:tautau}
\end{figure}

The $H\rightarrow \tau\tau$ decay is the most sensitive channel for leptonic Higgs boson decays, due to its high branching ratio. 
The most recent analysis uses the full Run~1 dataset with a centre-of-mass energy of $7\,\mathrm{TeV}$ and $8\,\mathrm{TeV}$ and investigates the ggF and VBF production mode~\cite{Aad:2015vsa}. All decays of the $\tau$ lepton (both leptonic, $\tau_{\mathrm{lep}}$, and hadronic, $\tau_{\mathrm{had}}$) are considered, leading to three analysis channels denoted $\tau_{\mathrm{lep}}\tau_{\mathrm{lep}}$ , $\tau_{\mathrm{lep}}\tau_{\mathrm{had}}$, and $\tau_{\mathrm{had}}\tau_{\mathrm{had}}$. Following a preselection, detailed in Ref.~\cite{Aad:2015vsa}, events in each channel are split into two categories. The VBF category contains events with two jets separated in pseudo-rapidity and targets signal events produced through VBF production. The ``Boosted'' category targets signal events where the Higgs boson has been produced with a large boost, primarily from ggF production, and requires the transverse momentum of the reconstructed Higgs boson candidate to be greater than 100 GeV. 
Combining analysis channels and categories yields six signal regions. Separate BDTs are trained for each, using between six and nine input variables. The selection of these variables has been separately optimised for each signal region, and chosen in order to exploit discriminating features such as resonance properties, event activity and topology, as well as the characteristic VBF topology in the corresponding category. One of the most important input variables is the mass of the di-$\tau$ system, the reconstruction of which is quite challenging due to the presence of at least two neutrinos in the final state; the Missing Mass Calculator (MMC) is used for this purpose~\cite{Elagin:2010aw}. Figure~\ref{fig:tautau}(a) shows a comparison of reconstructed masses for the data-based estimation of $Z\rightarrow \tau\tau$ (see next paragraph) and signal Monte Carlo in the $\tau_{\mathrm{lep}}\tau_{\mathrm{had}}$ channel. Good discrimination is observed.

The background composition differs between each analysis channel, but in all three the most important background is the irreducible $Z\rightarrow\tau\tau$ production. This background is modelled using the ``embedding'' technique, where $Z\rightarrow\mu\mu$ events are selected in data and the reconstructed muons are replaced by simulated $\tau$ lepton decays. Thus, the $Z$ boson kinematics and all other event activity comes entirely from data. Another important background in the $\tau_{\mathrm{lep}}\tau_{\mathrm{had}}$ and $\tau_{\mathrm{had}}\tau_{\mathrm{had}}$ channels stems from jets being mis-identified as hadronic $\tau$ leptons. Fully data-driven techniques are used for the estimation of backgrounds from misidentified $\tau$ decay products, described in detail in Ref.~\cite{Aad:2015vsa}. In addition, control regions are introduced to normalise the background contributions from top-quark production and $Z\rightarrow \ell\ell$ production. 

The signal strength and cross section are extracted by fitting the BDT score with signal and background templates simultaneously in the six SR together with nine control regions to better constrain the background normalisations. 
An excess of data events over the background prediction is observed in all three sub-channels.
The observed signal strength is $\mu = 1.43^{+0.31}_{-0.29}\mathrm{(stat.)} ^{+0.41}_{-0.30}\mathrm{(syst.)}$ for a Higgs boson mass of $125\,\mathrm{GeV}$.
Figure~\ref{fig:tautau}(b) shows the observed value of $\mu$ in the individual channels. Leading systematic uncertainties are the theoretical systematic uncertainty on the ggF production, the uncertainty on the normalisation of the $Z\rightarrow\ell\ell$ and top quark backgrounds in $\tau_{\mathrm{lep}}\tau_{\mathrm{had}}$, and the uncertainty on the jet energy scale calibration.
The observed (expected) significance of the measurement is $4.1\sigma$ ($3.2\sigma$)  providing direct evidence for $H\rightarrow \tau\tau$ decays.

\section{Measurements in $H \rightarrow \mu\mu$}

Searches for the $H \rightarrow \mu\mu$ decay are performed to establish the Yukawa coupling of the Higgs boson to second generation fermions. While the mass resolution of the  $H \rightarrow \mu\mu$ channel is good yielding a clean experimental signature, the sensitivity of the search is severely restricted by the low branching ratio of this decay for the Higgs boson in the SM which is $2.18\times10^{-4}$~\cite{deFlorian:2016spz}. The search is performed investigating the ggF and VBF Higgs boson production mode using $36.1\,\mathrm{fb}^{-1}$ of Run~2 $pp$ collision data at $13\,\mathrm{TeV}$~\cite{Aaboud:2017ojs}.

Events are selected for the analysis if they contain exactly two isolated muons of opposite charge with $p_{\mathrm{T}}> 15\,\mathrm{GeV}$. To ensure a high trigger efficiency, the leading muon must have $p_{\mathrm{T}}> 27\,\mathrm{GeV}$. Jets are required to have $p_{\mathrm{T}}\,{>}\,25\,\mathrm{GeV}$ if $|\,\eta\,|\,{<}\,2.4$ and $p_{\mathrm{T}}\,{>}\,30\,\mathrm{GeV}$ for $2.4\,{<}\,|\,\eta\,|\,{<}\,4.5$. To reduce the background from top-quark production, a $b$-jet veto is applied and events are required to have
$E_{\rm T}^{\rm miss}< 80\,\mathrm{GeV}$. The search is performed by fitting the dimuon invariant mass distribution, $m_{\mu\mu}$, which is required to be $110 < m_{\mu\mu} < 160\,\mathrm{GeV}$ in the signal region. This relatively wide range is chosen to allow determination of the background shape and normalisation from the sidebands.

\begin{figure}[bt]
\centering
\includegraphics[height=2.5in]{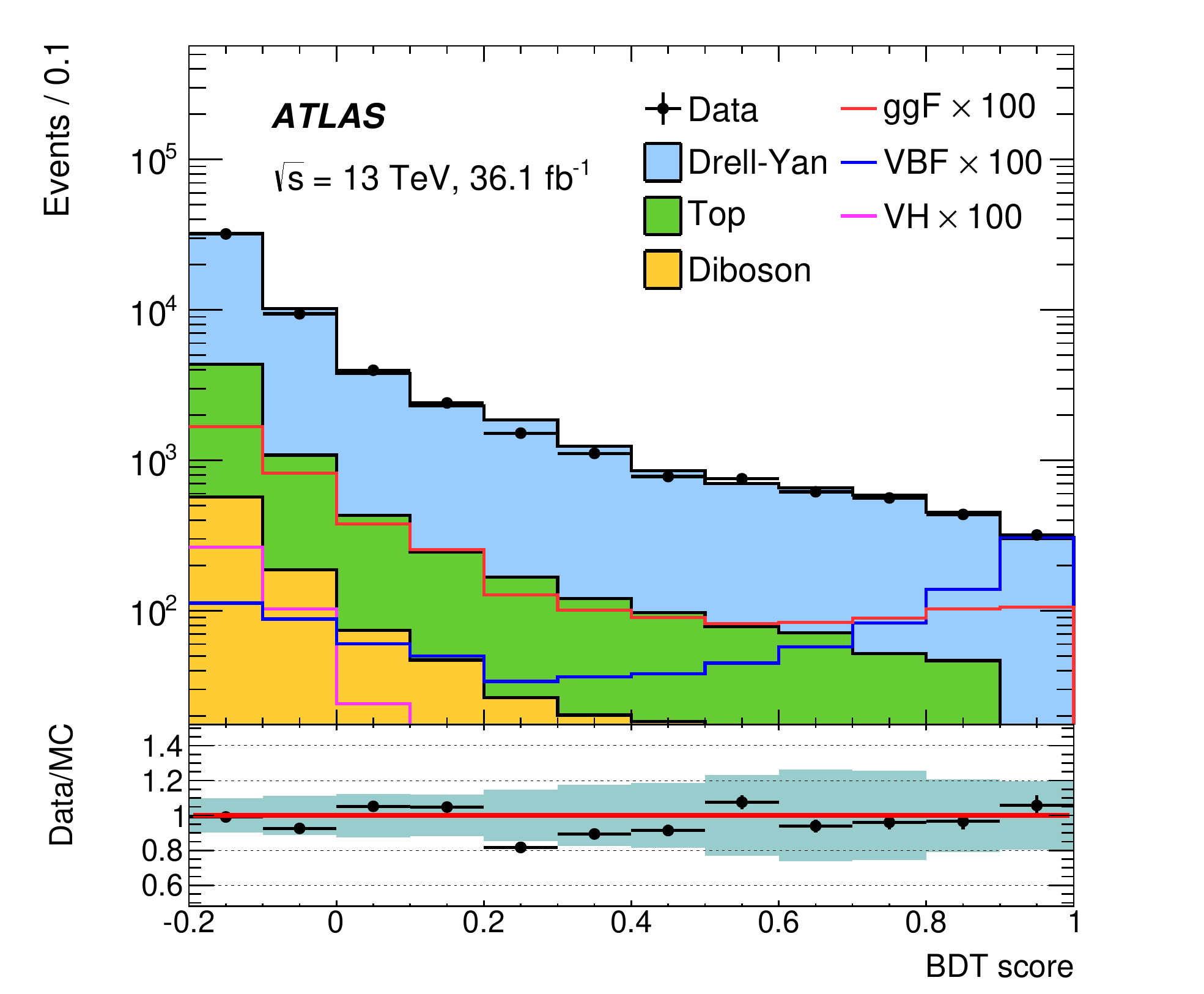}
\hspace{8mm}
\includegraphics[height=2.5in]{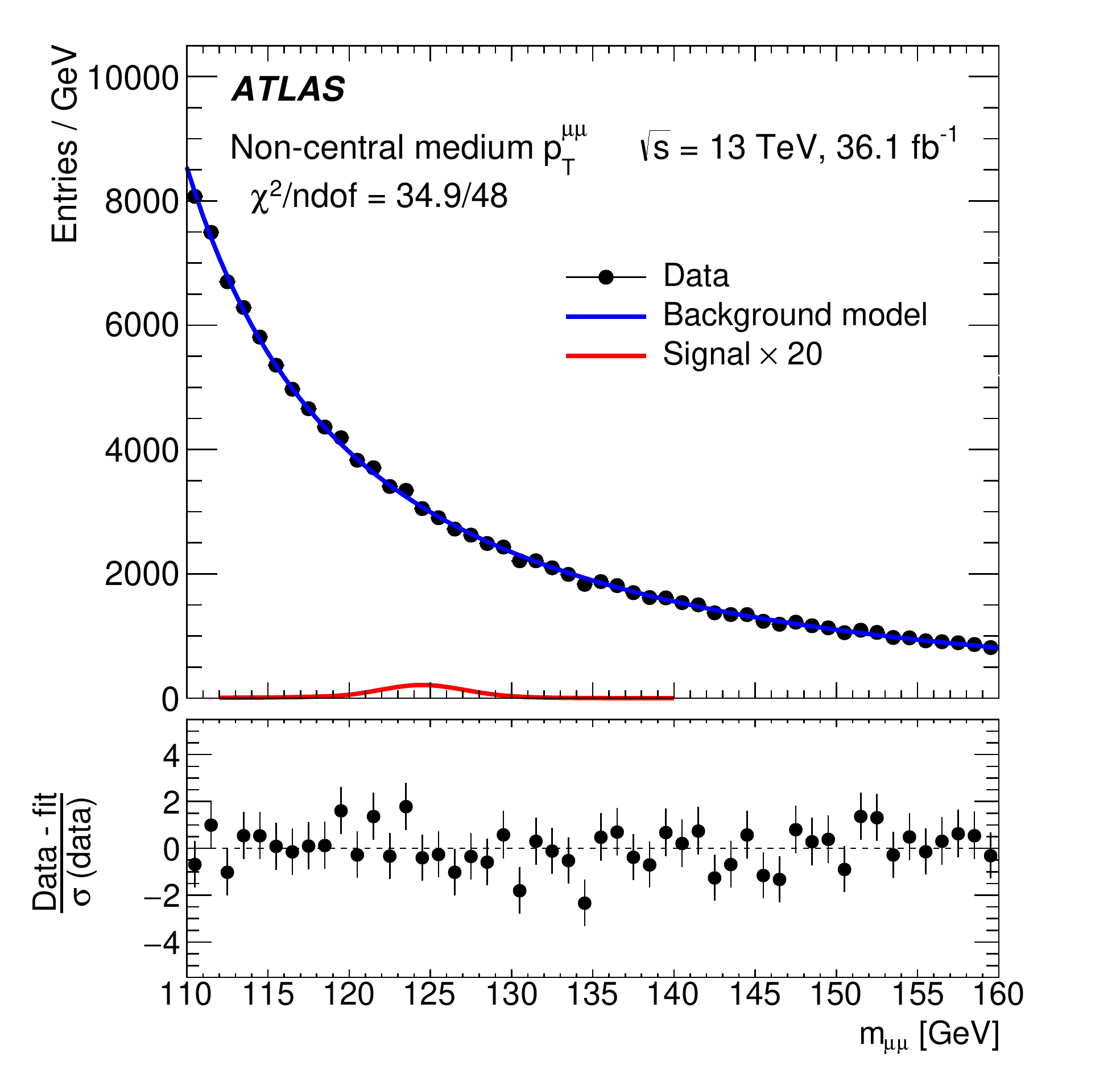}
\caption{ Fig.~(a) shows the observed and simulated distributions of the BDT score for events in the inclusive signal region for $H \rightarrow \mu\mu$ that contain at least two jets~\cite{Aaboud:2017ojs}. The error band only reflects the statistical and experimental uncertainties in the MC background prediction, while the theoretical uncertainties are not included. Fig.~(b) shows the background-only fit to the observed $m_{\mu\mu}$ distributions in the non-central medium $p_{\mathrm{T}}^{\mu\mu}$ category~\cite{Aaboud:2017ojs}. Only the statistical uncertainties are shown for the data points. The expected signal is scaled by a factor of 20.}
\label{fig:mumu}
\end{figure}

The signal region is then divided into six categories targeting ggF production and two categories targeting VBF production. 
The VBF categories are only considered for events containing at least two jets. Here, a BDT is trained using di-muon and di-jet variables as well as muon-jet variables. Figure~\ref{fig:mumu}(a) shows the BDT score. The two VBF categories correspond to the bins $0.7 < \textrm{BDT score} < 0.9$ and $\textrm{BDT score} \ge 0.9$. The remaining events are then split into categories according to the transverse momentum of the dimuon system ( $p_{\mathrm{T}}^{\mu\mu}$): low ($< 15\,\mathrm{GeV}$), medium
($15 - 50\,\mathrm{GeV}$) and high ($> 50\,\mathrm{GeV}$). Events in each of these three categories are then further classified into a central (if both muons have $|\eta| < 1$) and a non-central (if one or both muons have $|\eta| > 1$) category. The reason for this final classification is that muon momentum resolution in the barrel region is better, thus producing a narrower invariant mass distribution and a higher overall sensitivity. 

Analytic models are used to describe the signal and background  distributions, detailed in Ref.~\cite{Aaboud:2017ojs}. The signal is described by the sum of a Gaussian and a Crystal Ball function; the parameters of the model are obtained from fits to simulated SM Higgs boson samples.
The adopted model for the background is the sum of a Breit-Wigner function convolved with a Gaussian function, and an exponential function divided by a cubic function.
The results of the search are obtained by a binned maximum-likelihood fit of the sum of the signal and background models to the observed $m_{\mu\mu}$ distributions in the range 110-$160 \,\mathrm{GeV}$. The fit is done simultaneously in all categories. Figure~\ref{fig:mumu}(b) shows the background-only fit to the observed $m_{\mu\mu}$ distribution in the non-central medium $p_{\mathrm{T}}^{\mu\mu}$ category.

The resulting signal strength is $\mu = -0.1 \pm 1.5$. Thus, no excess is seen over the background expectation and an exclusion limit can be set on the $H\rightarrow\mu\mu$ signal strength. For a Higgs boson mass of $125\,\mathrm{GeV}$
the observed (expected) limit at 95\% confidence level (CL) is 3.0 (3.1) times the SM prediction.
This limit is driven by the data statistical uncertainty. When combined with
the ATLAS Run 1 data, the observed (expected) upper limit is 2.8 (2.9) at the 95\% CL.
 
\section{Conclusions}

A review of the latest ATLAS results in measurements of  Higgs boson decays to $WW^{\ast}$, $\tau\tau$, and $\mu\mu$ has been presented. 
Observation of the Higgs boson decay to $WW^{\ast}$ is established with the analysis of Run~1 data. Additionally a first look at the VBF and $WH$ production modes in Run 2 has been presented. Both analyses are still statistically limited. Evidence of the Higgs boson decay to $\tau\tau$ has been established with the analysis of Run~1 data. The measured signal strength of $\mu = 1.43^{+0.31}_{-0.29}\mathrm{(stat.)} ^{+0.41}_{-0.30}\mathrm{(syst.)}$ is in agreement with the SM prediction. A search for $H\rightarrow\mu\mu$ decays has been performed with the 2015+2016 Run 2 dataset and combined with the Run 1 dataset. The upper limit at the 95\% CL is 2.8 times the SM prediction.



\begin{thebibliography}{99}


\bibitem{Aad:2008zzm}
  ATLAS Collaboration,
  JINST 3 (2008) S08003.

\bibitem{ATLAS:2014aga}
  ATLAS Collaboration,
  Phys.\ Rev.\ D\ 92, 012006 (2015)
  [arXiv:1412.2641 [hep-ex]].

\bibitem{Aad:2016lvc}
  ATLAS Collaboration,
  JHEP 08 (2016) 104
  [arXiv:1604.02997 [hep-ex]]

\bibitem{Aad:2015ona}
  ATLAS Collaboration,
  JHEP 08 (2015) 137
  [arXiv:1506.06641 [hep-ex]].

\bibitem{HWWConf} 
  ATLAS Collaboration,
  ATLAS-CONF-2016-112,
  http://cds.cern.ch/record/2231811.

\bibitem{cart84} 
  L. Breiman, J. Friedman, R. Olshen and C. Stone,
  Wadsworth and Brooks, 1984.
 
\bibitem{FREUND1997119} 
  Y. Freund, R. E. Schapire,
  J. Comput. Syst. Sci. 55 (1997) 119.  
 
\bibitem{Friedman2002367} 
  J. H. Friedman,
  Comput. Stat. Data Anal. 38 (2002) 367Ð378.
  
\bibitem{Aad:2015vsa}
  ATLAS Collaboration,
  JHEP 04 (2015) 117
  [arXiv:1501.04943 [hep-ex]].

\bibitem{Elagin:2010aw}
  A.~Elagin, P.~Murat, A.~Pranko and A.~Sofanov,
  Nucl. Instrum. Meth. A654 (2011) 481Ð489
  [arXiv:1012.4686 [hep-ex]].


\bibitem{deFlorian:2016spz}
  LHC Higgs Cross Section Working Group.
   [arXiv:1610.07922 [hep-ph]].


\bibitem{Aaboud:2017ojs}
  ATLAS Collaboration,
  Phys.\ Rev.\ Lett.\ 119 (2017) 051802
  [arXiv:1705.04582 [hep-ex]].
   
  
%
%



\end{thebibliography}
\end{document}